\input harvmac
\overfullrule=0pt
\def\Title#1#2{\rightline{#1}\ifx\answ\bigans\nopagenumbers\pageno0\vskip1in
\else\pageno1\vskip.8in\fi \centerline{\titlefont #2}\vskip .5in}

\font\ticp=cmcsc10
\font\secfont=cmcsc10

%
%
\baselineskip=18pt plus 2pt minus 2pt

\def\ajou#1&#2(#3){\ \sl#1\bf#2\rm(19#3)}
%
\def\CH{{\cal H}}

%

\def\z{ z}
%


\def\r{\rho}

\def\s{\sigma}
\def\g{\gamma}
\def\t{\tau}
\def\a{\alpha}
\def\b{\beta}
\def\d{\delta}

\def\p{\pi}

\def\O{\Omega}

\def\l{{\lambda}}

\def\G{\Gamma}
\def\L{\Lambda}

%


\def\TrH#1{ {\raise -.5em
                      \hbox{$\buildrel {\textstyle  {\rm Tr } }\over
{\scriptscriptstyle \CH _ {#1}}$}~}}
\def\char#1#2{\vartheta{\bigl[ {#1\atop  #2}\bigr] }}
\def\charz#1#2{Z{\bigl[ {#1\atop  #2}\bigr] }}
\def\charbig#1#2{\vartheta{\biggl[ {#1\atop  #2}\biggr] }}

\def\detF{D_F{\bigl[ {a\atop b}\bigr] }}
\def\bZ{{\bf Z}}
\def\IZ{\relax\ifmmode\mathchoice
{\hbox{\cmss Z\kern-.4em Z}}{\hbox{\cmss Z\kern-.4em Z}}
{\lower.9pt\hbox{\cmsss Z\kern-.4em Z}}
{\lower1.2pt\hbox{\cmsss Z\kern-.4em Z}}\else{\cmss Z\kern-.4em Z}\fi}
\def\IC{\relax\hbox{$\inbar\kern-.3em{\rm C}$}}
\def\IR{\relax{\rm I\kern-.18em R}}
\def\1{\relax 1 { \rm \kern-.35em I}}
\font\cmss=cmss10 \font\cmsss=cmss10 at 7pt

%

\def\frac#1#2{{#1 \over #2}}

\def\p+{{\partial_+}}

\def\half{{1 \over 2}}

\def\dbar{{\bar d}}

\def\grad{\bigtriangledown}

\def\apm{\alpha^{\prime}}


%

\Title{\vbox{\baselineskip12pt
\hbox{CALT-68-2051}
\hbox{\ticp doe research and}
\hbox{\ticp development report}
\hbox{}
\hbox{hep-th/9604178}
}}
{\vbox{\centerline {\bf STRINGS ON ORIENTIFOLDS} }}

\centerline{{\ticp
Atish Dabholkar\footnote{$^1$}{e-mail: atish@theory.caltech.edu}
and Jaemo Park\footnote{$^2$}{e-mail: jpk@theory.caltech.edu}
}}

\vskip.1in
\centerline{\it Lauritsen Laboratory of  High Energy Physics}
\centerline{\it California Institute of Technology}
\centerline{\it Pasadena, CA 91125, USA}

\vskip .1in

\bigskip
\centerline{ABSTRACT}
\medskip

We construct several examples of compactification of Type IIB
theory on orientifolds and discuss their duals. In six dimensions
we obtain models with $N=1$ supersymmetry, multiple tensor
multiplets, and different gauge groups. In nine dimensions we
obtain a model that is dual to M-theory  compactified on a Klein
bottle.

\bigskip

\bigskip
\Date{April, 1996}

\vfill\eject

\def\npb#1#2#3{{\sl Nucl. Phys.} {\bf B#1} (#2) #3}
\def\plb#1#2#3{{\sl Phys. Lett.} {\bf B#1} (#2) #3}
\def\prl#1#2#3{{\sl Phys. Rev. Lett. }{\bf #1} (#2) #3}
\def\prd#1#2#3{{\sl Phys. Rev. }{\bf D#1} (#2) #3}

\def\cmp#1#2#3{{\sl Comm. Math. Phys. }{\bf #1} (#2) #3}
\def\mpl#1#2#3{{\sl Mod. Phys. Lett. }{\bf #1} (#2) #3}

%

\lref\DLP{
J. Dai, R. G. Leigh, and J. Polchinski,
\mpl{A4}{1989}{2073}\semi
R. G. Leigh, \mpl{A4}{1989}{2767}. }

\lref\Sagn{
A. Sagnotti, in Cargese '87, ``Non-perturbative Quantum
Field Theory,'' ed. G. Mack et. al. (Pergamon Press, 1988) p. 521;
{\it Some Properties of Open-String
Theories,} preprint ROM2F-95/18, hep-th/9509080\semi
M. Bianchi and A. Sagnotti,
\plb{247}{1990}{517}; \npb{361}{1991}{519}}

\lref\Hora{
P. Horava, \npb{327}{1989}{461};  \plb{231}{1989}{251};
\plb{289}{1992}{293}; \npb{418}{1994}{571}.}

\lref\Sen{A. Sen, ``M-Theory on $(K3 \times S^1 )/Z_2$,'' MRI-PHY/07/96,
hep-th/9602010.}

\lref\GiPo{E.~G.~Gimon and J.~Polchinski, ``Consistency Conditions
for Orientifolds and D-manifolds,'' hep-th/9601038.}

\lref\Polc{
J.~Polchinski, \prl{75}{1995}{4724}, hep-th/9510017.}

\lref\Erle{J.~Erler, ``Anomaly Cancellation in Six Dimensions,''
{\sl J. Math. Phys.}{\bf 35}(1988) 377.}

\lref\GrSc{
M. B. Green and J. H. Schwarz, \plb{149}{1984}{117}\semi
\plb{151}{1985}{21}.}

\lref\PoCa{
J. Polchinski and Y. Cai, \npb{296}{88}{91}.}

\lref\CLNY{
C. G. Callan, C. Lovelace, C. R. Nappi and S.A. Yost,
\npb{308}{1988}{221}.}

\lref\EGH{T. Eguchi, P.~B.~Gilkey, and A.~J.~Hanson,
{\sl Phys. Rept.} {\bf 66} (1980) 213.}

\lref\CLNYII{
C. G. Callan, C. Lovelace, C. R. Nappi and S.A. Yost,
Nucl. Phys. {\bf B293}, 83 (1987).}

\lref\GSWII{M. B. Green, J. H. Schwarz, and E. Witten,
{\it Superstring Theory},  {\rm Vol. II} ,
Cambridge University Press (1987).}

\lref\AGWi{L. Alvarez-Gaum\'e and E. Witten,
\npb{234}{83}{269}.}

\lref\Schw{ J.~H.~Schwarz, \plb{371}{1996}{223}, hep-th/9512053.}

\lref\SagnI{A. Sagnotti, \plb{294}{1992}{196}.}

\lref\Vafa{C.~Vafa, ``Evidence for F-theory,'' HUPT-96/A004,
hep-th/9602022.}

\lref\Witt{E.~Witten, ``Five-branes and M-Theory on an Orbifold,''
hep-th/9512219.}

\lref\WittII{E. Witten, ``Some Comments on String Dynamics,''
hep-th/9507121}

\lref\Mumf{D. Mumford, {\it Tata Lectures on Theta {\rm I}}, Birkh{\"a}user
 (1983).}

\lref\AGMV{L. Alvarez-Gaum{\' e}, G. Moore and C.  Vafa,
\cmp{106} {1986}{40}.}

\lref\CHL{S.~Chaudhuri, G.~Hockney, J. Lykken, \prl{75}{1995}{2264}.}

\lref\ChPo{S.~Chaudhuri and J. Polchinski, \prd{52}{1995}{7168}.}

\lref\HoWi{P.~Horava and E.~Witten, \npb{460}{1996}{506}.}

\lref\PCJ{J.~Polchinski, S.~Chaudhuri, C.~V.~Johnson, ``Notes
on D-Branes,'' NSF-ITP-96-003, hep-th/9602052.}
\lref\Mumf{D. Mumford, {\it Tata Lectures on Theta {\rm I}}, Birkh{\"a}user
 (1983).}
\lref\AGMV{L. Alvarez-Gaum{\' e}, G. Moore and C.  Vafa,
\cmp{106} {1986}{40}.}

\lref\DaPa{A.~Dabholkar and J.~Park, ``An Orientifold of Type IIB theory
on K3,'' CALT-68-2038, hep-th/9602030.}

\lref\SeWi{N. ~Seiberg and E.~Witten, ``Comments on String Dynamics in
Six Dimensions,'' RU-96-12, IASSNS-HEP-96/19,  hep-th/9603003.}

\lref\KuRa{A. Kumar and K. Ray, ``M-Theory on Orientifolds of
$K_3 \times S^1$,'' hep-th/9602144.}

\lref\VaMo{C.~Vafa and D.~Morrison, ``Compactifications of F-Theory on
 Calabi-Yau Threefolds-I, II,'' hep-th/9602114, hep-th/9603161.}

\lref\GiJo{E. Gimon and C. Johnson, ``K3 Orientifolds,'' hep-th/9604129.}
\lref\Sen{A. Sen, ``M-Theory on $(K3 \times S^1)/Z_2$,'' hep-th/9602010 ;
``Orbifolds of M-Theory and String Theory,'' hep-th/9603113.}

\lref\VaWi{C.~Vafa and E.~Witten, ``Dual String Pairs with  $ N=1$ and $ N=2$
Supersymmetry in Four Dimensions,'' HUTP-95-A23, hep-th/9507050.}

\lref\Mobi{This possibility has been considered in  informal discussions  but
we are not certain of its origin.}

\lref\Walt{M.~A.~Walton, \prd{37}{1988}{377}.}
\newsec{Introduction}

In this paper we discuss string compactifications on orientifolds
to six and higher dimensions.
Orientifolds are a generalization of  orbifolds \refs{\Polc,  \DLP, \Sagn,
\Hora}
in which
the orbifold symmetry includes orientation reversal on the
worldsheet (for a  review see  \PCJ\  and references therein).
Orientifolding allows one  to construct
new perturbative vacua that cannot be obtained by usual
Calabi-Yau compactification of  string theory.  One can thus explore
different regions in the moduli space of string vacua that  were previously
not accessible.

In six dimensions we focus on orientifolds of Type IIB theory compactified
on a $K3$ orbifold to obtain six dimensional theories  with $N=1$
spacetime supersymmetry.  It has recently become clear that
the dynamics  of  $D=6$, $N=1$ string theories is quite rich and offers many
surprises.
There are points in the moduli spaces of these theories where tensionless
strings appear which  makes it possible to have non-trivial dynamics in
the infra-red \refs{\SeWi, \Witt}.
In particular, there can be phase transitions  in which
the number of tensor multiplets can change.  It is therefore quite interesting
to analyze different branches of the tensor-multiplet moduli space.
Usual Calabi-Yau compactifications
can give only one tensor multiplet.  In \DaPa\ an orientifold
was constructed  that has nine tensor multiplets. In this paper we discuss
some generalizations that give  models with  five, seven, nine, or ten tensor
multiplets
with different gauge groups.   Models with multiple tensor-multiplets
can also be obtained by compactifications of M-theory
\refs{\SeWi, \Sen,  \KuRa},
or of F-theory \refs{\Vafa, \VaMo}.
The orientifolds  that we construct allow  one to study the duals
of  some of these compactifications as perturbative string theories.

In nine dimensions we consider an orientifold of Type IIB theory compactified
on
a circle to obtain  the dual of
M-theory compactified on a Klein bottle.  It is interesting
to note that the  compactification of M-theory on a circle gives the Type IIA
theory, on an interval the $E_8\times E_8$ heterotic string \HoWi, on a
M\"obius
strip a CHL string\refs{\CHL, \ChPo},  and on a torus the Type II string \Schw.
Thus, compactification on a Klein bottle completes this list of Ricci-flat
compactifications to nine and ten dimensions.
We also discuss some issues regarding the compactification of Type I theory
on a torus.

This paper is organized as follows. In section two we first
discuss some generalities about orientifolds.
In section three we discuss orientifolds  of  toroidal compactifications.
In section four we discuss orientifolds of Type IIB theory compactified on K3
orbifolds.
The calculation of tadpoles and the relevant partition sums are summarized
in the Appendix.

\newsec{Some Generalities about Orientifolds}

In general our starting point will be  some $\bZ_N$ orbifold
of toroidally compactified Type IIB theory.  We can
then take the orientifold projection $(1 +\Omega \b)/2$, where
$\Omega$ is the orientation reversal on the worldsheet and
$\b$ is some $\bZ_2$ involution of the orbifold.
If the orbifold group $\bZ_N$  is generated by the element $\a$,
then the total projection we would like to perform is given by
$(\frac{1+\a +...+ \a^{N-1}}{N})(\frac{1+\Omega \b}{2})$
in both the twisted and the untwisted sectors of the orbifold.
The orientifold group $G$  can be written as
$G=G_1 +\Omega G_2$ such that $\Omega h \Omega h'  \in G_1
\, {\rm for}\, h, h' \in G_2$.

The closed string sector of the orientifold is obtained by
projecting  the spectrum of the original orbifold onto  states
that are invariant under the orientifold symmetry.
The open-string sector of  the orientifold arises as follows.
Orientifolding introduces unoriented surfaces
in the closed-string perturbation theory.
The unoriented surfaces such as the Klein bottle
can have tadpoles of R-R fields in the closed string tree
channel. The tadpoles correspond to the fact that the equations
of motion for some  R-R fields are  not satisfied because the orientifold
plane acts as the source of the R-R fields \Polc.
By including
the right number of D-branes which are also sources for the R-R fields
with opposite charge,  one can cancel these tadpoles.
This introduces the open-string sector with
appropriate boundary conditions and Chan-Paton factors.
As we shall see, sometimes the Klein bottle amplitude
turns out to have no tadpoles;  in these cases there is no
need to introduce the open-string sector, and the closed-string
sector by itself describes a consistent theory.

An open string can begin on a D-brane labeled by $i$ and end
on one labeled by $j$. The label of the D-brane is the
Chan-Paton factor at each end.
Let us denote a  general state in the open string sector
by $|\psi, ij \rangle$. An element of $G_1$ then acts
on this state as
\eqn\gone{
g:\qquad |\psi,ij\rangle \ \rightarrow
(\gamma_g )_{ii'}|g\cdot\psi,i'j'\rangle
(\gamma_g^{-1})_{j'j},}
for some unitary matrix $\gamma_g$ corresponding to $g$.
Similarly, an element of $\Omega G_2$ acts as
\eqn\gtwo{
\Omega h:\qquad |\psi,ij\rangle \ \rightarrow
(\gamma_{\Omega h})_{ii'}|\Omega h\cdot\psi,j'i'\rangle
(\gamma_{\Omega h}^{-1})_{j'j}.}

The relevant partition sums for the Klein bottle,
the M\"obius strip, and the cylinder  are respectively
$\int_0^{\infty}dt/2t$ times
\eqn\traces{\eqalign{
{\rm KB:} &\quad {\rm Tr}_{\rm NSNS + RR}^{\rm U+T}
\left\{ \frac{\Omega \b}{2}\,\frac{1+ \a+...+\a^{N-1}}{N}\,\frac{1+(-1)^{F}}{2}
e^{-2\pi t(L_0 + \tilde L_0)}\right\} \cr
{\rm MS:} &\quad {\rm Tr}_{\rm NS-R}^{\l\l}
\left\{ \frac{\Omega \b}{2}\,\frac{1+\a+...+\a^{N-1}}{N}\,\frac{1+(-1)^{F}}{2}
e^{-2\pi t L_0}\right\} \cr
{\rm C:} &\quad {\rm Tr}_{\rm NS-R}^{\l\l'}
\left\{ \frac{1}{2}\,\frac{1+\a+...+\a^{N-1}}{N}\,\frac{1+(-1)^{F}}{2}
e^{-2\pi t L_0}\right\} .\cr }}
Here $F$ is the worldsheet fermion number,
and as usual $\frac{1+(-1)^{F}}{2}$ performs the GSO projection.
The Klein bottle includes contributions both from the
untwisted sector(U) and the twisted sectors(T) of the original
orbifold.  Orientation reversal $ \Omega$ takes NS-R sector
to R-NS sector, so these sectors do not contribute to the
trace. The labels $ \l$
and $ \l'$ refer to the type of  D-brane an open string ends on.
For example, in a theory with both 5-branes and 9-branes,  $ \l$ and $\l'$
are either $ 5$ or $ 9$; one has to include the sectors $ 55$ and
$ 99$ for the M\"obius strip, and the sectors $ 55$, $ 99$, $ 59$, and $ 95$
for the cylinder\GiPo.
The  tadpoles can be extracted by  factorizing  the
loop-amplitude in the tree channel.
Tadpole cancellation  then determines the number of D-branes
as well as the form of the $\gamma$ matrices introduced
earlier, which in turn determines the open string sector completely.
In fact in many examples
that we consider,  spacetime supersymmetry and anomaly cancellation
usually place powerful  constraints which determine the spectrum
even without knowing the full  form of the $\gamma$ matrices.

Many of the details of the tadpole calculation are similar to
those discussed in \refs{\PCJ, \DaPa, \GiPo} and will not be
repeated here.
We give a  collection of relevant partition sums and their factorized forms
in the tree channel in the Appendix.

\newsec{Orientifolds of Toroidally Compactified Type IIB theory.}

\subsec{An Example in Nine Dimensions}

Consider Type IIB theory compactified say in the
$X^9$ direction  on a circle $S_9$ of radius $r_9$. We can take an orientifold
with the group $\{ 1, S\Omega\}$ where  $S$ is a half-shift
along the circle, $X^9 \rightarrow X^9 + \pi r_9$.
The closed-string sector of this theory is obtained by projecting
onto  states that are invariant under $S\Omega$.
The massless bosonic spectrum of  Type IIB theory in ten dimensions consists
of  the metric $g_{MN}$, the dilaton $\phi^1$, and a  two-form $B^2_{MN}$
from the NS-NS sector;   a two-form $B^1_{MN}$, a scalar
$\phi^2$, and a four-form
$A_{MNPQ}$  with self-dual field strength from the R-R
sector. The fields $g_{MN}$, $\phi^1$, and $B^1_{MN}$ are all
even under $\Omega$, whereas the fields
$A_{MNPQ}$, $B^2_{MN}$, and  $ \phi^2$ are odd.  If we were projecting
only under $\Omega$, we would obtain the spectrum of Type I
strings; the superscript $1$ above refers to the fields that survive this
projection.

Now,  if we expand a given field $\Psi$  in terms of the Kaluza-Klein momentum
modes $\Psi_m$ carrying quantized momentum $m/R$ then the
modes with even $m$ are even under $S$, whereas the modes with odd
$m$ are odd. Thus, the combined projection under $\Omega S$
eliminates all odd momentum modes of the fields $g_{MN}$, $\phi^1$, and
$B^1_{MN}$,
but all even momentum modes of $A_{MNPQ}$, $B^2_{MN}$, and $ \phi^2$.
In particular,   once we restrict ourselves to zero momentum
modes to obtain the massless spectrum  in nine dimensions,
we obtain the closed string sector
of  the Type I string reduced to nine dimensions.

Let us now look at the open-string sector. As explained in the previous
section, open-string sector arises from the addition of D-branes
to cancel tadpoles in the Klein bottle amplitude. Now, because of the
half-shift that accompanies $\Omega$, only states with odd
winding appear in the crosscap state and are thus massive.
Another way to see this is to first compute the amplitude in the loop channel
and then factorize in the tree channel. The loop channel momentum
sum gives a term proportional to $\sum_m (-1)^m e^{\frac{-t \apm{m^2}}{r_9^2}}$
where $t$ is the loop-channel parameter.
To see the tadpoles in the tree channel we  use Poisson resummation
formula and take the limit $t \rightarrow 0$ corresponding to long, thin tubes;
it is easy to see that in this limit the amplitude vanishes, and there is no
tadpole.
Therefore,  to  obtain a consistent orientifold there is no need to add any
branes.

To see what this theory is dual to,  we compactify further on a circle $S_8$
of radius $ r_8$ in the direction $ X_8$.
The Type IIB theory is  T-dual to Type IIA under $ r_8 \rightarrow 1/r_8$,
and moreover the operation $\Omega$ in IIB is dual to $R_8 \Omega$
in IIA where $R_8$ is the reflection $X_8 \rightarrow -X_8$ \PCJ.
Now  Type IIA theory is  M-theory compactified on a circle $S_{10}$
in the $ X^{10}$ direction.  The operation $ R_8\O$
 corresponds, in M-theory, to taking $X^{8}\rightarrow -X^8$, at the same
time flipping the sign of the three-form potential  $C_{MNP}$ of the eleven
dimensional supergravity. In M-theory we can interchange the two circles $ S^8$
and $ S^{10}$.
Therefore, the combined operation
$S\Omega$ in Type IIB theory
 correponds, in M-theory, to
$X^{10} \rightarrow  -X^{10} $,
$X^9 \rightarrow  X^9 + \pi r_9$ which
 is nothing but the $Z_2$ transformation that turns  the torus $ T_{9, 10}$
into a Klein bottle.
Notice that this is not a purely geometric operation in M-theory but is
accompanied by a simultaneous change of sign of the three-form
potential.
Under  the interchange of  the two circles $ S_{10}$ and $ S_{8}$,
the symmetry $ R_8 \Omega $ in Type IIA theory is conjugate
to the  symmetry $(-1)^{F_L}$, where
$F_L$ is the spacetime fermion number coming from the left-movers
\VaWi.
All R-R fields are odd under this symmetry and all NS-NS fields are even.
Thus, the strong coupling limit of  the orbifold of Type-IIA theory
under the combined operation  $ (-1)^{F_L}$ and $X_9 \rightarrow X_9 + \pi r_9
$
is given by M-theory compactified on a Klein bottle.

It is amusing that we have an example of
a  compactification on a non-orientable surface.
Another example is M-theory on a M\"obius strip
which is dual to a CHL compactification
\refs{\CHL, \ChPo}.  Recall that the $E_8\times E_8$ string is dual
to M-theory on an interval in the tenth direction: the two
$E_8$ factors  live  at the two endpoints of the interval \HoWi.
Compactifying further on a circle, we obtain M-theory on a cylinder.
The CHL string is obtained
as a $Z_2$ orbifold of the heterotic string in nine dimensions.
The orbifold symmetry corresponds to an  interchange of the two $E_8$
factors  accompanied by a half shift on the circle. The combined operation
is again $X^{10} \rightarrow  - X^{10}  $, $X^9 \rightarrow X^9 + \pi r_9 $
which turns the cylinder into a M\"obius strip \Mobi.

\subsec{Type I Theory in Eight Dimensions}

Type I theory compactified in  the $8$ and $ 9$ directions
to eight dimensions can be viewed
as an orientifold of the Type IIB theory on the torus $T_{89}$.
It is straightforward to find the massless spectrum,  but there
is one subtlety in taking the T-dual of this theory which is worth
mentioning.

Let us T-dualize first in the $X^9$ direction. T-duality is a one
sided  parity transform \PCJ\ which means that in  the RNS formulation
of the superstring, only the left-moving coordinate ${\tilde X}^9$ and its
fermionic partner ${\tilde \Psi}^9$  change sign.  Thus, T-duality takes
Type IIB theory to Type IIA theory,  and takes $\Omega$ to
$R_9 \Omega$, where $R_9$ is the reflection in the $X^9$ direction.
If we dualize again in the $X^8$ direction,
we would get Type IIB theory back;
$\Omega$  goes to $R_{89} \Omega$, where $R_{89}$ reflects
both $X^8$ and $ X^9$.
This identification leads to the following puzzle for the  orientifold
with the group $\{ 1, R_{89} \Omega\}$.  Under $\Omega$ the
four-form field $A_{MNPQ}$ is odd, therefore the modes like
$A_{MNP9}$ and $A_{MNP8}$ which are 3-forms in eight dimensions
would be even under the combined
operation $R_{89} \Omega $ and would survive the projection.
But $N=1$ supersymmetry in $D=8$ uniquely determines the massless
field content and does not allow a three-form potential. Therefore,
supersymmetry
is broken by this projection. On the other hand, the orientifold
with the group $\{ 1, R_{89} \Omega\}$ is T-dual
to the one with the group $\{ 1,  \Omega\}$, and we cannot break supersymmetry
by a T-duality transformation.
We should really have  obtained the T-dual of
Type I strings in eight dimensions.  The reason for this discrepancy is
that Type IIB theory has an additional symmetry $(-1)^{F_L}$ under
which all R-R fields are odd.  The correct projection that gives the T-dual
of Type I theory involves the combined operation $R_{89} (-1)^{F_L} $ instead
of
just the geometric reflection.

It is easy to see this ambiguity on the worldsheet.
In the Ramond sector,  the zero modes
${\tilde \Psi}^{M}$ correspond to the $\Gamma^M$ matrices of the spacetime
Clifford algebra.  Under the  T-duality transformation ${\tilde \Psi}^9
\rightarrow
-{\tilde \Psi}^9$,  the spinors transform as
\eqn\spinor{\eqalign{
S &\rightarrow S\cr
{\tilde S} &\rightarrow \Gamma^9 \Gamma{\tilde S},\cr
}}
where $S$ and ${\tilde S}$ are the right-moving and
left-moving spacetime spinors respectively, and $\Gamma$ as usual is
the matrix that anticommutes with all $\Gamma^M$ matrices and squares to one.
If we T-dualize further in the $X^8$ direction then
$S$ goes to itself,  and $ {\tilde S}$ goes to
$ \Gamma^8 \Gamma \Gamma^9 \Gamma{\tilde S} =  \Gamma^9\Gamma^8
{\tilde S}$. Let us now see how the massless fields from the Ramond-Ramond
sector transform. The vertex operator for  an n-form field strength
$H_{M_1...M_n}$
is proportional to ${\bar S} \Gamma_{M_1...M_n} {\tilde S}$ where
$  \Gamma^{M_1...M_n} = \frac{1}{n!} (\Gamma^{M_1}...\Gamma^{M_n}
\pm {\rm permutations})$.
It  is easy to see that the effect of T-duality on the R-R
field strengths $H_{M_1...M_n}$ and the corresponding potentials
is to remove the $8, 9$ indices if they are present and add them if they are
not.
For example,  the vertex operator for $H^{1}_{M89}$ is proportional
to ${\bar S} \Gamma_{M} \G_8\G_9 {\tilde S}$. Under
T-duality, it would map onto ${\bar S} \Gamma_{M} {\tilde S}$
which is the vertex operator for the field strength of a scalar.
Thus, $B^1_{89}$  maps onto  the scalar $\phi^{2}$.
However, because $\Gamma^8 \Gamma$ and $\Gamma^9 \Gamma$
anticommute with each other, there is a choice of sign for  the
action on the R-R fields, which corresponds precisely to the choice
between $R_{89}$ and $R_{89} (-1)^{F_L}$. This ambiguity
is, of course, fixed by the correct choice of the orientifold symmetry.

\newsec{Orientifolds of Type IIB theory on $K3$}

\subsec{General Remarks}

Let us review some relevant facts about  the  $K3$ surfaces which can
be represented  as $\bZ_N$ orbifolds of the 4-torus $T^4$ \Walt.  Let
$(z_1, z_2)$ be the complex co-ordinates on the torus, and consider the
$\bZ_N$ transformation  generated by
\eqn\trans{
g: \quad (z_1, z_2) \rightarrow (e^{2\pi i/N} z_1,  e^{-2\pi i/N}z_2).
}
The $ \bZ_N$ group must be a subgroup of $ SU(2)$ to obtain
unbroken supersymmetry in six dimensions.
The torus $T^4$ is obtained by identifying a lattice $\L$ of points
in $R^4$, so the orbifold group must leave the lattice invariant
to have a sensible action on the torus. This crystallographic condition
allows only four possibilities:
the groups  $\bZ_2$ and $\bZ_4$ when $\L$ is the square ($SU(2)^4$)  lattice
given by  the identifications $z_i \sim z_k +1,   \sim z_k + i,  k =1,2 $;
or $\bZ_3$ and $\bZ_6$  when $\L$ is the hexagonal  ($SU(3)^2$) lattice given
by the identifications $z_i \sim z_k  +1,  \sim z_k  + e^{2\pi i/3} ,  k =1,
2$.
At a fixed point of a  $\bZ_k$ symmetry there is a curvature singularity.
A smooth $K3$ can be obtained by blowing  up the singularity by
replacing a ball around the  fixed point by  an appropriate
smooth non-compact Ricci-flat surfaces $E_k$ whose boundary
at infinity is $S^3/{\bZ_k}$.

In this section we consider two classes of orientifold projections
$(1 + \Omega \b )/2$ of Type IIB theory
on these orbifolds. In the first  class of models
 we take $\b$ to be identity, whereas
in the second class  we take
$\b$ to be a specific $\bZ_2$ involution $S$ of $K3$ that has $8$ fixed
points. We shall  give an explicit description of this involution
in the following subsections.

One immediate question is whether the projection leaves any supersymmetries
unbroken. In the case of $\Omega$ the combination $Q_{\a} + \Omega {\tilde
Q}_{\a}$
of the left-moving and right-moving supercharges will be invariant;
supersymmetry
will be broken by half, giving us $N=1$ supersymmetry starting from $N=2$.
When we combine $\Omega$ with  $S$, we do not want to break the supersymmetry
further, so $S$ should leave all $N=2$ supersymmetries invariant.
This is possible if the rotational part of the symmetry $S$
is a subgroup of $SU(2)$, or equivalently if it leaves the holomorphic
2-form invariant.
It is useful to consider the example  of $\bZ_2$ orbifold.  In this case we
have
$\a : (z_1,  z_2) \rightarrow (-z_1, -z_2)$ which generates
 a discrete subgroup  of the $SU(2)$ holonomy group of a smooth
$K3$, and therefore leaves two supercharges invariant giving us $N=2$
supersymmetry.
The  symmetry $S$ is given by
$S: (z_1, z_2) \rightarrow (-z_1 + \half , -z_2 + \half)$ which is
a combination of a shift and a rotation \DaPa.  The shift has no effect on the
supercharges;
the rotation is again a subgroup of the holonomy group $SU(2)$ and therefore
does not
break any supersymmetries by itself. Thus the combined operation $S\Omega$
gives  $N=1$ supersymmetry as required.
Now, the $\bZ_2$ orbifold  admits other involutions; for example,
the Enriques involution $E: (z_1, z_2) \rightarrow (-z_1 + \half ,   z_2 +
\half)$
which does not leave the holomorphic 2-form invariant, and cannot
be used for orientifolding if we want unbroken supersymmetry.

The closed-string sector of an orientifold can be
determined  by index theory and by appropriate projection.
Recall that  the massless representations in $D=6$ are labeled by the
representations
of the little group  which is $Spin(4)\sim SU(2) \times SU(2)$.
The massless $N=1$ supermultiplets are

1. the gravity multiplet: $(3, 3) + (1, 3)  + 2(2, 3) $,

2. the vector multiplet:   $(2, 2) + 2 (1, 2)$

3. the tensor multiplet:  $(3, 1) + (1, 1) + 2(2, 1)$

4. the hyper multiplet:  $4(1, 1) + 2(2, 1)$.

To determine the massless modes we need to know  the Dolbeault cohomology
\GSWII, and how the symmetry $ \O \b$ acts on the cohomology.
For a smooth  $K3$,  the nonzero Hodge numbers are $h^{00}=h^{22}=
h^{02}=h^{20}=1$, and $h^{11}=20$.  Among the 2-forms  the $(0, 2)$,
$(2, 0)$, and the K\"ahler $(1, 1)$ form are self-dual, and the remaining
$19$ $(1, 1)$ forms are anti-self-dual.
The  manifolds $E_k$ have $(k-1)$ anti-self-dual
$(1, 1)$ harmonic forms, and one $(0, 0)$ form. In the orbifold limit, each
fixed point
that is repaired by $E_k$ contributes $(k-1)$ anti-self-dual $(1, 1)$ forms
which together with the  $(1, 1)$ forms of the original torus that
are invariant under the orbifold group give the $20$ $(1, 1)$ forms
of $K3$.

It is useful to think in terms of  Type I theory compactified on a smooth $
K3$.
In this case, the orientation reversal symmetry in ten dimensions, which
we shall call $ \O_0$ has the effect of flipping the sign of $ A_{MNPQ}$,
$ \phi^2$, and $ B^2_{MN}$, leaving other massless fields invariant.
The resulting theory has $ h^{11}(= 20)$ hypermultiplets which come
from  the zero modes of $ B^{1}_{MN}$ and $ g_{MN}$. There is only one
tensor multiplet from contracting $ B^{1}_{MN}$ with the $ (0, 0)$ form.
Now imagine performing a projection not with $ \O_0$ but with $\O_0 T$ where
$ T $ is some geometric symmetry under which $n_T$ $ (1, 1)$ forms are
odd and all others are even. In this case,
by contracting $ A_{MNPQ}$ with these $(1, 1)$ forms,
one can obtain $ n_T$ additional
tensor multiplets that are invariant under the combined operation
$ \O_0 T$ .
At the same time, $ n_T$ hyper-multiplets are now projected out
changing their total number to  $ (20 -n_T)$.
This reasoning gives the simple equation
\eqn\tensor{
n_T + n^c_H = 20,}
where $n^c_H$ refers to the number of hypermultiplets arising
from the closed string sector, and $n_T+1$  is the total number
of tensor multiplets. Moreover,   no vector multiplets arise from the closed
string sector because there are no harmonic odd forms on
$K3$, so starting with even forms and the metric in ten dimensions,
one cannot obtain a one-form vector potential.
We can thus read off the closed string spectrum  immediately from the
 geometric data of the orientifold.

In the orbifold limit,  the orientifold symmetry $ \Omega$, for the purposes
of counting of states, is really a combination of $ \Omega_0 T$ where
$ T$ is some geometric symmetry that has nontrivial action
on the cohomology. This is because at each fixed point, $ \Omega$ takes the
sector twisted by $ \r$ to the one twisted by $ \r^{-1}$.
If we repair the singularity at the fixed point of a $\bZ_k$ symmetry by the
smooth
surfaces $ E_k$ then the $ (k-1)$
$\left( 1, 1 \right)$-forms  coming from  $E_k$ correspond to the $ (k-1)$
twisted sectors. If we think of the orbifold as a limit of a
smooth $ K_3$,  then except in the case when $ \a$ is  a $ \bZ_2$  twist,  we
get
a nontrivial action on the cohomology denoted by $T$.
This information is sufficient to work out the spectrum of the orientifold
in the closed-string sector.

Let us now discuss the massless bosonic spectrum coming from
the NS open-string sector. The states
\eqn\vectors{
\psi_{-1/2}^\mu|0,ij\rangle \lambda_{ji}, \quad \mu = 1, 2, 3, 4,}
belong to the vector multiplets whereas the states
\eqn\scalars{\psi_{-1/2}^m|0,ij\rangle \lambda_{ji}, \quad  m= 6, 7, 8, 9,}
belong to the hypermultiplets.
We have to keep only the states that are
invariant under $\a$ and $\Omega \b$.  For this purpose we need
to know the form of the $ \g$ matrices defined in \gone\ and \gtwo\
which are determined by the requirement of tadpole cancelation.
The  Chan-Paton wave functions $\lambda_{ij}$
allowed by these projections determine the gauge group and the
matter representations.

There are some features of the tadpole calculation that are common
to all orbifolds. First, by the arguments given in \GiPo, only 5-branes
and 9-branes  appear.  Let $ v_6$ and $ v_4$ be the regularized volumes
of the noncompact and the compact spaces in string units.  If we look at the
the Klein bottle amplitude in the tree channel then non-zero tadpoles
proportional
to $ v_6 v_4$ correspond to 10-form exchange requiring addition of 9-branes.
Similarly a term proportional to $ v_6/ v_4$ corresponds to the exchange of
6-forms from the untwisted sector, requiring addition of 5-branes, and the
terms
proportional to $ v_6$ correspond to the exchange of  6-forms from
the twisted sector and must cancel without the addition of any branes.
Now with  the orientifold group $ G= G_1 + \Omega G_2$, 9-branes can
arise only if $G_2$ contains the identity, and 5-branes arise only if
$G_2$ contains the element $ R$ that reflects all four internal co-ordinates.
In these cases the determination of the 10-form and the untwisted 6-form
tadpoles is identical to the calculation in \GiPo\ which requires  $ 32$
9-branes with $ \g_{\O, 9}^T =\g_{\O, 9}$, and/or  $32$ 5-branes with
$ \g_{\O, 5}^T = -\g_{\O, 5}$.

\subsec{${\bZ_2} $Orbifold}

For the $\bZ_2$ orbifold, the model in the first class with the projection
$(1+\O)/2$ has been discussed
in \GiPo,  and the model in the second class with the
projection $(1+S\O)/2$ in \DaPa. We would now like
to consider a model that is closely related to the one in \DaPa.
Let us recall that in  \DaPa\ the symmetry $S$ was chosen to be
such that  $S^2=1$. However, if we are on a $ \bZ_2$ orbifold,
then the symmetry
can square to  the element $ \a$ that generates the  orbifold group.
We choose
\eqn\stwo{
S: \quad (z_1,  z_2) \rightarrow (iz_1, -iz_2).}
Now $S$ has $4$ fixed points and not $8$. However,
they are also the fixed points of $\a$ which is a $\bZ_2$
symmetry. So on the orbifold, the fixed point of $S$ should
be regarded as having Euler character $2$ giving us the total Euler character
of $8$ in agreement with the Lefschetz number \EGH.

Obviously,  the spectrum consists of
the closed string sector found in \DaPa\
giving us $n_T =8$,  $n_H =12$ and the  gravity multiplet.
However, because now neither $ R$ nor the identity are
elements of $ G_2$,  there is no need to add  any branes,
and there is no open-string sector. One nontrivial check
is that the tadpoles  of the
R-R fields from the twisted sector
now have to cancel by themselves  for the Klein bottle without any
contribution from the open-string sector.
It is easy to see using the formulae in the Appendix that the tadpoles
from the untwisted sector  cancel against those from
 the sector twisted by $\half$ giving us a consistent theory.
Gravitational anomalies cancel  completely as expected.

\subsec{{$\bZ_3$} Orbifold}

The orbifold symmetry in this case  has nine fixed points of
order $3$ which contribute two anti-self-dual (1, 1)
forms each  giving $18$ in all.  Out of the six  2-forms on
the torus one  anti-self-dual (1, 1) form and the remaining
three self-dual 2-forms  are invariant under
$\a$ giving us $22$ 2-forms of the $K3$.

Let us first consider the projection under $\Omega$.
As explained in ${\S 4.1}$,  at each fixed point of the
orbifold $\Omega$ interchanges the  sector twisted
by $\a$ to that twisted by $\a^{-1}$ besides flipping
the sign of all R-R fields.  This means that of the two  tensor
multiplets coming from each fixed point,  only one will be
invariant,  giving us $n_T=9$ from the nine fixed points,
and $n^c_H = 11$ from \tensor .

To determine the open-string sector we note that,
by the general arguments mentioned in  {\S 4.1},
there will be $ 32$  9-branes, and we can choose
$\gamma_{\O}= {\bf 1}$ by a unitary change of basis \GiPo.
The requirement that $ ({\O \a})^2 = {\a}^2$ implies
\eqn\gthree{
\g_{\a^2} =\g_{\a}^2= \g_{\O\a} (\g_{\O\a}^{-1})^T.}
Using the fact the the $\g$ matrices are unitary, and $\g_{\O \a } = \g_{\O}
\g_{\a}$, we conclude
that  $\g_{\a}$ is real. Furthermore, because $ \g_{\a}^3 =1$,
the only eigenvalues are cube-roots of unity. If   $ n$ eigenvalues
are $e^{2\pi i/3}$,  then  $ n$ will be $e^{-2\pi i/3}$,
and $ 32-2n$ will be $ 1$.  We can then write $ \g$ in a block-diagonal
form where in a $ 2n$ dimensional subspace it acts as a $ 2\pi/3 $ rotation
and in $ 32-2n$ dimensional subspace it  equals the identity matrix. This
information and  anomaly cancellation is enough to determine that
$ n=8$.  We can also verify this  by a detailed calculation of tadpoles
as  discussed in the Appendix.
The gauge group will then be given by  $ SO(16) \times U(8)$
with hypermultiplets in  $ (1, 28) + (16, 8)$.
It is easy to see that  the anomaly terms
proportional to $ {\rm tr}(F^4)$ and $ {\rm tr}(R^4) $ vanish.
It is not necessary for the remaining anomaly to factorize because
we have more than one tensor multiplet  available,  and the anomalies
can be canceled by the generalized Green-Schwarz mechanism as
in \refs{\GrSc, \SagnI, \DaPa}.

Let us now describe the action of $S$ on the $ \bZ_3$ orbifold.
It is given by
\eqn\sthree{
S: \quad (\z_1, z_2) \rightarrow (-z_1, -z_2).}
$ S$ has $ 16$ fixed points on the torus but on the orbifold they
split into one singlet and five triplets of $ \bZ_3$.  The Euler character
of the fixed point
at the origin which is a singlet under the $ \bZ_3$ is
$ 3$ and that of  the $ 5$ triplets is $ 1$ each  giving  $ 8$ altogether.
Now,  because $ S$ is just a reflection of all co-ordinates,  the orientifold
with the projection $ (1+ S\Omega)/2$ is T-dual to the one  described in
the previous paragraphs with the projection$ ( 1 + \Omega)/2$.
T-duality turns 9-branes into 5-branes, but the spectrum
remains unchanged.

\subsec{{$\bZ_4$} Orbifold}

The $ \bZ_4$ orbifold has four fixed points of order $ 4$. Each contributes
three tensor multiplets out of which only one is invariant under the action $
\O$.
No additional tensors arise from the  six  doublets of fixed points of order $
2$.
Altogether $ n_T= 4$, and $ n^c_H = 16$.
In this case both 5-branes and 9-branes will be present,
and we can choose
\eqn\some{
\g_{\O , 9} = {\bf 1} , \quad
 \g_{\O , 5} = {\bf J} \equiv \left [\matrix{
0 & -i \cr
i & 0 \cr
}\right ].}
The remaining algebra is determined in terms of
the matrices $ \g_{\a , 9}$ and $ \g_{\a ,  5}$.
Tadpoles are canceled if
$ {\rm Tr} \left(  \g_{\a ,  9}\right) = {\rm Tr} \left(  \g_{\a ,  9}\right)^2
 ={\rm Tr} \left(  \g_{\a ,  9}\right)^3 =0 $ and similarly for the
matrices with subscript $ 5$. This determines the $ \g$
matrices completely.  Moreover  $ \g_{\a ,  9} = \g_{\a ,  5}$, and
their eigenvalues
are such that  each forth root of unity appears eight times.
The gauge group is
$ U\left( 8 \right) \times U\left( 8 \right) \times U\left( 8 \right) \times
U\left( 8 \right) $
with hypermultiplets in
$ \left( 28, 1, 1, 1 \right) + \left( 1, 28, 1, 1 \right) +
\left( 1, 1, 28, 1 \right) + \left(1, 1, 1, 28 \right) +
\left( 8, 8, 1, 1 \right) + \left( 1, 1, 8, 8 \right) + \left( 8, 1,  8,
1\right)
+ \left( 1, 8, 1, 8 \right)$.
Once again the anomaly terms proportional to $ {\rm tr} \left( F ^4\right)$
for each factor, and the coefficient of ${\rm tr}\left( R^4\right)$ vanish.

Let us now consider the action of the symmetry $ S$
which is given by
\eqn\sfour{
S: \quad (\z_1, z_2) \rightarrow (-z_1 + \frac{1+i}{2}, -z_2  +\frac{1+i}{2}).}
This form is determined by the requirement that $ S$ has to preserve the
orbifold symmetries; in particular, it should map a fixed point of a given
order to a fixed point of the same order. It is easy to check that
eight  $ (1, 1)$ forms are odd under $ S$. The $ 16$ fixed points form
four  quartets under $ \bZ_4$. In addition, $ S$ leaves two  doublets under
$ \a$ invariant which should be regarded as fixed points on $ K_3$ with
Euler character $ 2$.  The total Euler character of the fixed point set
 adds up to $ 8$.

If we consider the orientifold with the projection
$ \left( 1 + \Omega S \right)$, then only  $ 32$ 5-branes are required.
As in \DaPa\ we find $ n_T = 8 $, $ n^c_H =12$ from the closed-string
sector. We can  place $ 16$  branes at
a  fixed point of $ \a^2$ which is  in a doublet  of $ \a$ that is  left
invariant by
$ S$, and $ 16$ at its image under $ \a$.  For example, we can place $ 16$
branes at the  $ \left( \half, \half \right)$ and the remaining $ 16$ at
$ \left( \frac{i}{2}, \frac{i}{2} \right)$. In this case
the gauge group is $ U(8) \times U\left( 8 \right)$,
with charged hyper-multiplets in $ 2 \left( 8, 8\right)$.
This is exactly the spectrum of the model considered in \DaPa\  for the $\bZ_2$
orbifold. If we place $ 16$ branes at the fixed point of $ \a$,
and $ 16$ at its image under $ S$, then the gauge group is
$ U(4) \times U(4) \times U(4) \times U(4)$ with hypermultiplets
in $ (4, 4, 1, 1) + (4,  1, 4, 1) + (1, 4, 1, 4) + (1, 1, 4, 4)$.

\subsec{{$\bZ_6$} Orbifold}

In this case,  we get two tensors from the fixed
points of order $ 6$ and one each from the four fixed
points of order $ 3$  giving us $ n_T = 6$ and $ n^c_H = 14$.
The open-string sector has   both 5-branes and 9-branes.
The eigenvalues of the matrix $ \g_{\a , 5} =  \g_{\a , 9}$ are as follows:
$ 1$ and $ -1$ appear eight times each and the other sixth roots
of unity appear four times each.
The resulting  gauge-group  is $U(4) \times U(4) \times U(8) $
with hypermultiplets in $ \left( 6, 1, 1 \right) +  \left( 1, 6, 1 \right) +
\left( 4, 1, 8 \right)
+ \left( 1, 4, 8 \right)$ from the $ 55$ sector, and identical spectrum from
the $ 99$ sector.  The $ 59$ sector contributes  hypermultiplets in
$ (4, 1, 1, 4, 1, 1) + (1, 4, 1, 1, 4, 1) + (1, 1, 8, 1, 1, 8)$.

\appendix{A}{Tadpole Calculation}

For evaluating the traces in the loop-channel we need the determinants
of chiral bosons and fermions with twisted boundary conditions.
Let us denote by  $\detF $
the fermion determinant of a chiral Dirac operator  $(\grad^z_{-\half})$
which corresponds to the  path integral of a complex chiral fermion with
boundary condition
$\psi (\s_1 +2\pi \, , \, \s_2) = -e^{2\pi i a} \psi (\s_1 \, , \, \s_2)$,
and $\psi (\s_1  \, , \, \s_2+2\pi) = -e^{2\pi i b} \psi (\s_1 \, , \, \s_2) $.
It is straightforward  to evaluate this determinant in the operator
formalism\AGMV .
Writing $q= e^{2\pi i \t}$,  and using the standard  relation between the path
integral and the operator formalism, it  is equal to the trace
$\TrH{}(h_b  \, q^ {H_a} )$.  $H_a$
is the Hamiltonian of a chiral, twisted fermion:
\eqn\hamiltonian{
H_a = \sum_{n=1} ^{\infty}(n-\half +a ) d_n^\dagger d_n +  (n-\half -a )
\dbar_n^\dagger \dbar_n
+\frac{a^2}{2} -\frac{1}{24}
}
The fermionic oscillators satisfy canonical anticommutation relations
 $\{ d_n^\dagger , d_m\} = \d_{mn}$ and $\{ \dbar_n^\dagger , \dbar_m\} =
\d_{mn}$, and
 $\CH $ is the usual Fock space representation of these commutations.
The group $Z_N$ acts on this Fock space through
$ h dh^{-1} = -e^{-2\pi i b} d$ , $ h \dbar h^{-1} = -e^{2\pi i b} \dbar$.
The trace equals (up to an arbitrary phase)
\eqn\trace{
e^{2\pi i a b} q^{\frac{a^2}{2}-\frac{1}{24}}
\prod_{n=1}^{\infty}  (1+q^{n-\half +a} e^{2\pi i b} )\,
 (1+q^{n-\half - a} e^{-2\pi i b}) \ .
}
Using the  product representation  of the theta function $\char{a}{b}(\t)$
with characteristics \Mumf\  ,   we see that
\eqn\detab{
\detF = \TrH{}(h_b  \, q^ {H_a} )
= \frac{\char{a}{b} (0|\t )} {\eta (\t )} \ ,
}
where  $\eta (\t )$ is the Dedekind $\eta  $ function.
The chiral boson determinant is
the inverse of the chiral fermion determinant,
except for $ a=\half$ when one needs to be careful about the zero modes.
Note that  untwisted NS fermions with
half-integer modings  and  antiperiodic boundary conditions for the trace
corresponds to $ a=0, b=0$;  an untwisted boson  with
periodic boundary condition along the $ \s_2$ direction
corresponds to $ a=\half, b=\half$.
Using these formulae one can write down the traces by inspection.
The tadpole calculation corresponding  to the 10-form and the  untwisted
6-form exchange  are identical to the one in \GiPo, and will not be repeated
here.
We shall be interested in the tadpole of only the 6-form from the twisted
sector which corresponds to the boundary conditions for  the
determinant for internal bosons  that have only oscillator sums
but  no momentum or winding sums.

Let us first evaluate the traces in \traces\ for the Klein bottle.
The total trace can be written as
\eqn\kbtrace{ \frac{(1-1)v_6}{64N} \int_0^{\infty} \frac{dt}{t^4}
8 \sum_{a, b}\charz{a}{b},}
where the $ (1-1)$ refers to NSNS - RR exchange in the tree channel,
$ v_6$ is $ V_6 /(4\pi \apm )^3$; $ b =k/N, k=1,..., (N-1)$ corresponding
to the terms with $ \a^k$ in the trace. Only the untwisted sector and the
sector twisted by $ \half $ contribute because for  other twisted sectors
$ \Omega$ is off-diagonal; $a$ is therefore either $ 0$, or $ \half$.
{}From the untwisted sector we get
\eqn\kbpartion{
\charz{0}{b} = 4 \sin^2 (2\pi b)
\frac{ \char{0}{\half}^2 \char{0}{2b + \half}  \char{0}{-2b -\half} }
{\eta^6 \char{\half}{2b + \half} \char{\half}{-2b - \half} };}
and from the sector twisted by $ \half$ at each fixed point that
is left invariant by $ \a^k$,  we get
\eqn\kbpartitiontwo{
\charz{\half}{b} =
-\frac{ \char{0}{\half}^2 \char{\half}{2b + \half}  \char{\half}{-2b -\half} }
{\eta^6 \char{0}{2b + \half} \char{0}{-2b - \half} }, }
where $ \t = 2 i t$ and $ b= k/N$.
Let us now turn to the traces for the cylinder. In this case, in general
we can have $ 55, 99, 59,$ or $ 95$ sectors. The partition sum
is given by
\eqn\ctrace{
\frac{(1-1)v_6}{64N} \int_0^{\infty} \frac{dt}{t^4}
 \sum_{\l, \l', b}\charz{\l \l'}{b} {\rm Tr} \, ({\g_{b, \l})
{\rm Tr} \, ({\g^{-1}_{ b, \l'}) }},
}
where $ \l$ and $ \l'$ take values either $ 5$ or $ 9$,
and ${\g_{\l, b} } $ refers to the matrix  ${\g_{\l, \a^k} } $ for
$ b= k/N$.
We obtain
\eqn\cpartition{\eqalign{
\charz{99}{b} =& \charz{55}{b} =  4 \sin^2 (\pi b)
\frac{ \char{0}{\half}^2 \char{0}{b + \half}  \char{0}{-b -\half} }
{\eta^6 \char{\half}{b + \half} \char{\half}{-b - \half} },  \cr
\charz{59}{b} =& \charz{95}{b} =
- \frac{ \char{0}{\half}^2 \char{\half}{b + \half}  \char{\half}{-b -\half} }
{\eta^6 \char{0}{b + \half} \char{0}{-b - \half} }, \cr
}}
with $ \t = it$.
The M\"obius strip amplitude  is given by
\eqn\mstrace{
\frac{(1-1)v_6}{64N} \int_0^{\infty} \frac{dt}{t^4}
 \sum_{\l,  b}\charz{\l \l}{b} {\rm Tr} \, ({\g^T_{b\O, \l}
{\g^{-1}_{b\O, \l}) }},
}
where only $ 55$ and $ 99$ sector contribute.  We obtain
\eqn\mspartition{
\charz{99}{b} =\,  \tan^2 (\pi b) \charz{55}{b} =  -4 \sin^2 (\pi b)
\frac{ \char{0}{\half}^2  \char{\half}{0}^2 \char{0}{b + \half}  \char{0}{-b
-\half}
\char{\half}{b }  \char{\half}{-b }}
{\eta^6 \char{0}{0}^2\char{\half}{b + \half} \char{\half}
{-b - \half} \char{0}{b }  \char{0}{-b }},
}
with $ \t = 2 i t$

To factorize in the tree channel we use the  modular transformations
 under $ \t \rightarrow -1/\t$:
\eqn\modularS{ \eqalign{
\charbig{a}{b}  (\t)\, =&  \,
(-i\t )^{-\half }e^{ -2 \pi i a b}  \charbig{-b}{a} (-1/\t) \cr
\eta (\t ) =& \left( -i \t  \right)^{-\half } \eta \left( -1/\t \right),\cr
}}
and take the limit $ t \rightarrow 0$.
While writing the tadpoles we also have to take into account
that the tree channel length  $ l $ is equal
to $ 1/4t$, $ 1/2t$, and $ 1/8t$ for the Klein bottle, the cylinder,
and the M\"obius strip respectively. The twisted-sector tadpole is
then proportional to  $ \frac{(1-1) v_6}{8N} \int dl $. In this common
normalization, we get,
\eqn\tadpole{\eqalign{
{\rm KB:} &\quad  (16)^2 \sin^2 (2\pi b), \qquad\qquad\,\,\,\,\qquad \qquad  a=
0, b \neq 0, \cr
        	    &\quad -64,             \qquad\qquad\qquad \,\,\,\,\,\qquad \quad
\qquad \quad a=\half  , b\neq 0, \half ;\cr
 {\rm C:}  &\quad 4 \sin^2 (\pi b)  {\rm Tr} \, (\g_{b, \l}){\rm Tr} \,
(\g^{-1}_{b, \l}),
\quad \,\,\, \,\qquad b\neq 0, \l =5 \, {\rm or }\, 9, \cr
&\quad   -{\rm Tr} \, (\g_{b, 5}){\rm Tr} \, (\g^{-1}_{b, 9}) - \left( 9
\leftrightarrow 5 \right),  \qquad b\neq 0;\cr
{\rm MS:}  &\quad  -64 \sin^2 (\pi b)  {\rm Tr} \, (\g^T_{b\O, 9}){\rm Tr} \,
(\g^{-1}_{b\O, 9}),  \quad b\neq 0, \half\cr
  &\quad  -64 \cos^2 (\pi b)  {\rm Tr} \, (\g^T_{b\O, 5}){\rm Tr} \,
(\g^{-1}_{b\O, 5}),
\,\,\,\,\, b\neq 0, \half. \cr
}}
The Klein bottle contributes  $ -64$ from each
sector twisted by $ \half$ for each fixed point
that is left invariant by $ \a^k$.

\bigbreak\bigskip\bigskip\leftline{{\secfont Acknowledgements}}\nobreak
This work was supported in part by the U. S. Department of Energy
under Grant No. DE-FG03-92-ER40701.
\bigskip\bigskip

\bigskip
\leftline{\bf Note Added:} In a paper \GiJo\ that appeared after this work
was completed  many of the orientifolds of $ K3$  discussed in \S 4 have been
found independently.

\vfill \eject
\listrefs
\end